# Portfolio-driven Resource Management for Transient Cloud Servers

PRATEEK SHARMA, DAVID IRWIN, and PRASHANT SHENOY, University of Massachusetts Amherst

Cloud providers have begun to offer their surplus capacity in the form of low-cost transient servers, which can be revoked unilaterally at any time. While the low cost of transient servers makes them attractive for a wide range of applications, such as data processing and scientific computing, failures due to server revocation can severely degrade application performance. Since different transient server types offer different cost and availability tradeoffs, we present the notion of server portfolios that is based on financial portfolio modeling. Server portfolios enable construction of an "optimal" mix of severs to meet an application's sensitivity to cost and revocation risk. We implement model-driven portfolios in a system called ExoSphere, and show how diverse applications can use portfolios and application-specific policies to gracefully handle transient servers. We show that ExoSphere enables widely-used parallel applications such as Spark, MPI, and BOINC to be made transiency-aware with modest effort. Our experiments show that allowing the applications to use suitable transiency-aware policies, ExoSphere is able to achieve 80% cost savings when compared to on-demand servers and greatly reduces revocation risk compared to existing approaches.

## 1 INTRODUCTION

Cloud computing has become popular in recent years for a wide range of applications, including latency-sensitive web services, computationally-intensive scientific workloads, and data-intensive parallel tasks. Due to their flexible resource allocation and billing model, cloud platforms are especially well-suited for running applications with dynamically varying workloads, or those that require compute resources for only short periods of time. Recently, cloud platforms have introduced a new class of servers, called *transient servers*, which they may unilaterally revoke at any time [46, 47]. Transient servers increase the utilization of cloud platforms, while enabling them to reclaim resources at any time for higher priority users.

Transient servers typically incur a fraction of the cost of their regular ("on-demand") server counterparts, making them a popular choice for running large-scale data-intensive jobs involving tens or hundreds of servers due to their low cost. However, revocations of some, or all, of an application's transient servers can seriously disrupt its performance or cause it to fail entirely. Thus, despite the low cost of transient servers, making effective use of this new class of server remains challenging.

On some cloud platforms, such as Amazon EC2, transient servers have dynamically varying prices that fluctuate continuously based on supply and demand. In addition, the availability of transient servers (in terms of their mean time to revocation), can also vary significantly across server configurations and based on changing market conditions. Unfortunately, cloud platforms generally do *not* directly expose availability statistics for transient servers, requiring users to indirectly infer them, e.g., via a price history or active probing [39]. Thus, it is challenging for a cloud application to judiciously select the most appropriate server configuration based on historical pricing or availability data to satisfy its needs. The problem is compounded by the large number of transient server configurations available to applications—there are over 2500 distinct types of transient servers in EC2 and over 300 in Google's cloud platform. Recent research suggests that mitigating revocation risk requires a parallel application to diversify across multiple transient server types, which further complicates decision making [43].

The preemptible nature of transient servers also imposes new requirements on cloud applications. Specifically, applications must determine whether and how to save their computation's intermediate state to gracefully handle server revocations, which are akin to server failures. Further, they must also define recovery policies to determine how to re-acquire new transient servers upon revocation, and how to restore state and resume their computation on these new servers. Different applications, such as Spark, MapReduce, and MPI, also have different tolerances to revocations, and require different application-specific mechanisms to handle revocations and their subsequent recovery. However, prior research has largely focused on separately designing custom modifications to support transiency for each narrow class of application [36, 43, 57].

To address this problem, we introduce a model-driven framework called *server portfolios*. Portfolios represent a virtual cloud cluster composed of a mix of transient server types with a configurable cost and availability depending on the application's tolerance to revocation risk and price sensitivity. Our portfolio model derives from Modern Portfolio Theory in financial economics [37, 38], which enables investors to methodically construct a financial portfolio from a large number of underlying assets with various risks and rewards.

The flexibility and explicit risk-awareness that portfolios offer is not provided by prior work on transient server selection. A majority of prior work [19, 25, 49, 59] on transient servers solves the problem of choosing *one* server type (among the hundreds that cloud providers offer). Choosing *multiple* server types has received relatively little attention, and mostly relies on application-specific, ad-hoc approaches to optimize either cost or revocation-risk [43]. In contrast, portfolios are a *general* technique that allow server selection for a wide range of risk tolerances and application preferences.

We use portfolio modeling as part of the design of an application-independent framework for supporting transiency, called ExoSphere. ExoSphere uses portfolio modeling to expose virtual clusters of transient servers of different types to different applications. Along with portfolio modeling, ExoSphere also supports custom application-specific policies for handling transiency. In particular, ExoSphere adopts an Exokernel approach [27] by exposing a set of basic mechanisms that are common to all transiency-enabled applications. These mechanisms can be used by applications to design policies for handling revocations, saving state, and performing recovery. Thus, ExoSphere's mechanisms simplify modifying distributed applications to effectively run on transient servers.

Although ExoSphere's design is general, its implementation builds on existing cluster managers, which often serve as the "operating system" for data centers and support multiple concurrent distributed applications, e.g., Spark and Hadoop. These systems, including Mesos [32] and Kubernetes [6], support a wide range of applications by exposing a core set of functions related to resource allocation, revocation, and scheduling. ExoSphere adapts and extends these functions to implement its portfolio abstraction and application-facing APIs. By using an existing cluster manager (Mesos), ExoSphere makes a useful practical contribution by extending it to run on transient cloud servers. In designing and implementing ExoSphere, this paper makes the following contributions:

**Portfolio Modeling**. We introduce a portfolio model for selecting transient cloud servers, and show how it can be used by different applications to create virtual clusters with configurable cost and revocation risk. Our portfolio model adapts and extends concepts from financial portfolio modeling to select a diverse set of servers for a distributed application. We show how portfolios allow applications to explicitly navigate the risk versus reward tradeoff.

**Risk Management Framework**. In addition to providing applications with model-driven portfolios, ExoSphere also allows applications to define transiency-specifc policies for managing server revocation risk. We show that these policies for handling transiency are inherently application-specific, and depend on a particular application's workload and risk tolerance. We distill a general set of transiency-specific policies and mechanisms into a well-defined API that it exposes to applications. Applications then implement various policies in handler functions defined in the API, which are triggered based on changing market conditions, e.g., when prices increase or a revocation occurs.

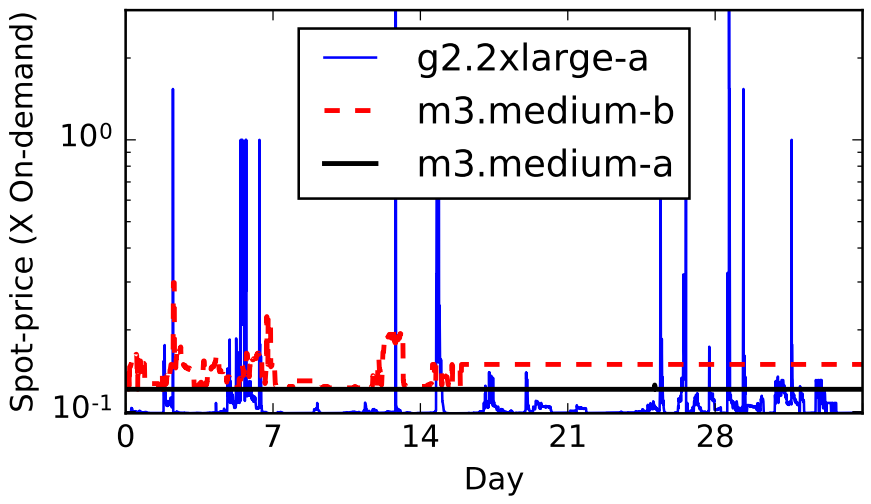

Fig. 1. EC2 spot prices for three servers over May 2015.

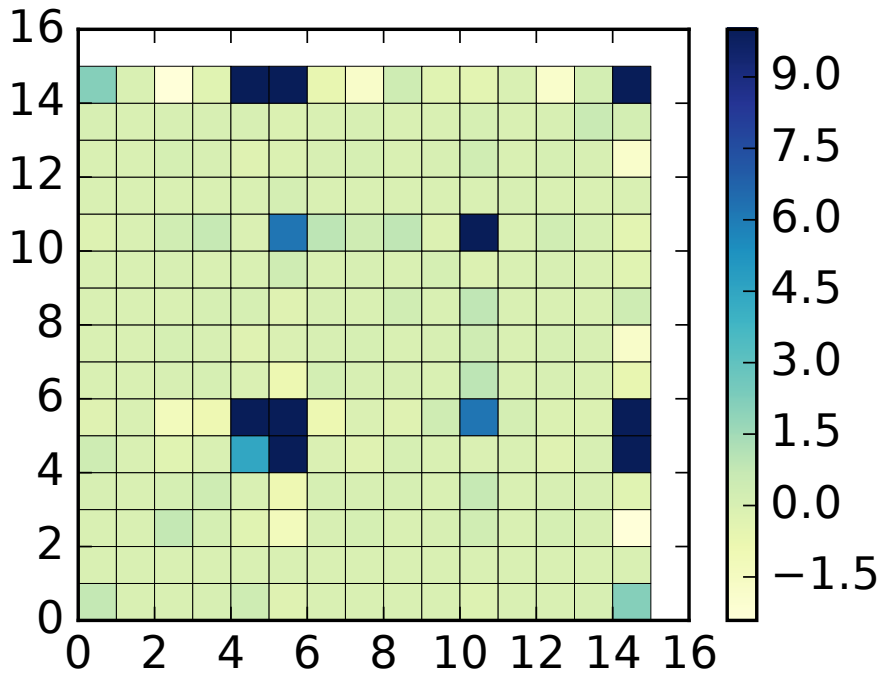

Fig. 2. Heat-map of covariances of spot prices of markets in the EC2 US-east-1 region. Significant number of markets are weakly and negatively correlated.

**Implementation and Evaluation**. Finally, we implement ExoSphere by extending Mesos, integrate it with EC2's spot market, and use it to implement transiency-aware variants of multiple applications. We show that ExoSphere simplifies the development of these applications, while also improving their performance and decreasing their costs compared to prior approaches. For example, we show that ExoSphere is able to achieve 80% reduction in cost compared to using on-demand servers with a negligible risk of revocation.

## 2 BACKGROUND & MOTIVATION

In this section, we provide background on transient cloud servers and applications capable of exploiting them.

**Transient Cloud Servers.** Traditional cloud servers are leased on an on-demand basis—cloud customers may request them when needed and the cloud platform provisions these servers until the customer relinquishes them. Since customer demand for cloud resources can be highly dynamic, the cloud platform needs to over-provision the aggregate server capacity to handle peak demand. Consequently, a significant portion of the cloud server capacity tends to be idle during non-peak periods. Cloud providers have begun to lease this surplus capacity at highly discounted prices to cost-sensitive customers. Doing so enables providers to earn revenue from otherwise idle resources. Such surplus servers are referred to as transient servers since the cloud provider can reclaim them from the customer at any time, e.g., when demand for standard on-demand servers begins to rise. Cloud providers typically provide a brief advance warning prior to preempting a transient server to enable the customer to gracefully shutdown the machine. The warning time currently ranges from two minutes in Amazon EC2 cloud to 30 seconds on Google's cloud platform.

Note that transient servers also arise internally in data centers, as cluster managers, such as Kubernetes, often include explicit support for background versus foreground tasks, such that background tasks are revoked if resources are needed for foreground tasks [6, 52]. While some prior work focuses on transient servers in this context, these internal dynamics are similar to the external dynamics of transient cloud servers, as they also arise from opportunistically leveraging idle resources. The key difference is that internal supply/demand dynamics are generally well-known by the data center, while the external supply/demand dynamics of transient cloud servers are only indirectly conveyed through price signals.

**Transient Server Pricing.** Different cloud providers have employed different approaches for pricing transient servers. Google's transient servers, called preemptible instances [5], offer a fixed 80% discount but also have a maximum lifetime of 24 hours (with the possibility of earlier preemption). In contrast, Amazon's transient servers (which are called spot instances [1]) offer a variable discount—the price of spot instances varies continuously based on market supply and demand for each server type (Figure 1). The customer specifies a maximum price (called a bid) that they are willing to pay when requesting spot servers. Amazon then runs a continuous sealed-bid multi-unit uniform price auction based off of the bids submitted, and determines a market price for the server [14]. If the market price increases above the bid price, then the server is immediately revoked after the warning period.

Since transient servers are surplus idle machines, the resources available in the transient server pool fluctuate continuously depending on the supply and demand of on-demand servers. Thus, whether a certain transient server is available depends on current market conditions. Furthermore, once allocated, the mean time to revocation may also vary over time. Importantly, cloud providers expose current prices for transient servers, but do not expose the surplus pool size or other availability metrics.

**Transient Server Applications.** Due to their preemptible nature, transient servers are typically not suitable for running interactive applications, such as web services, or any application that cannot tolerate downtime caused by server revocations. Batch-oriented disruption-tolerant applications are particularly well-suited for transient servers, since they can often tolerate longer completion times caused by occasional downtimes. A common scenario is to use tens or hundreds of transient servers to run highly parallel CPU-or-data-intensive applications at very low costs (when compared to standard on-demand server prices). However, if some fraction (or all) of the servers are revoked before the job has finished, the application has to resort to the following methods to resume computation.

The simplest approach (which is also the most wasteful) is to restart the entire job on new transient servers. Alternatively, if the data-intensive application supports checkpointing, then its state can also be checkpointed periodically and the job can be resumed from the most recent checkpoint after reacquiring lost servers. A final approach is to simply live migrate to new servers after receiving the revocation warning. However, migration is only feasible if an application's memory footprint is small enough to be completely copied before termination.

**Making Applications Transiency-aware.** To effectively use transient cloud servers to run parallel batch-oriented applications, we must address three key factors:

i) *Choosing the "right" server types*. Different server models offer different price and availability tradeoffs. Thus, server selection must consider the application's risk tolerance to revocation, as well servers' expected price, availability, and revocation frequency.

ii) *Choosing appropriate fault-tolerance mechanisms.* The revocation rates and delay-tolerance of the application influence the choice of fault-tolerance mechanisms such as checkpointing/restart, replication, or migration.

iii) *Recovering from server revocations.* If some fraction (or all) of an application's transient servers are revoked, the application must determine when and how many servers to reacquire and how to resume the computation via the chosen fault-tolerance mechanism.

In the rest of the paper, we present our model-driven portfolio framework for selecting transient servers and show how it can be used to run transiency-aware parallel applications. We also describe how ExoSphere can simplify the development of transiency-aware versions of applications, such as Spark, MPI, and BOINC with modest implementation overhead.

## 3 PORTFOLIO-DRIVEN SELECTION

A key factor in making effective use of transient servers is judiciously choosing the most appropriate server configuration for each application. Due to their preemptible nature and variable pricing, picking the "correct" server configuration is surprisingly complex in today's cloud platforms due to the following reasons:

**Large number of potential choices.** A typical cloud platform offers a large number of transient server markets. Amazon's EC2 cloud offers 2500 distinct markets, while Google Cloud Platform offers more than 300 markets for predefined machine types alone. Assuming an application imposes a certain base requirement on the desired per-server compute and memory capacity, it must still choose from a large number of feasible configurations.

**Pricing idiosyncrasies.** Cloud operators such as Amazon EC2 use demand-supply driven pricing to price their spot servers [10, 14]. Each server type has different demand-supply characteristics, and this can lead to some interesting idiosyncrasies, which can be seen in Figure 1. In this example, the `m3.medium` in availability zone `a` has the most stable prices, `g2.2xlarge` in the same availability zone has a lower average price but high variance, and the `m3.medium` in availability zone `b` has higher price than in zone `a`. The `g2.2xlarge` price spikes are not correlated with the other two servers. The example shows that smaller servers may occasionally be more heavily discounted than smaller servers, and that identical servers in two availability zones may also be discounted differently.

Importantly, choosing a server configuration based on price alone may yield sub-optimal results. For instance, server configurations with cheap prices may also see higher customer demand and consequently higher volatility and frequent revocations. Frequent revocations add substantial overheads to an application in terms of increasing checkpointing costs and adding recovery overheads. Instead, sometimes the choice of a slightly more expensive server configuration that sees a lower revocation rate may be a better choice and yield lower overall costs.

Revocation rates may also not be related to average prices—neither the willingness to pay higher prices (by using a higher bid) nor choosing higher priced configurations necessarily yield lower revocation rates [44]. It is not practical to expect applications to analyze detailed price histories and volatilities across hundreds of transient servers when choosing a server type.

Due to the challenges above, cloud providers such as Amazon have begun offering server selection tools. Amazon SpotFleet [3] automatically replaces revoked servers. However, SpotFleet provides a limited choice in terms of the combinations of server configurations it offers, and does not alleviate many of the challenges above. While it enables applications to specify their combination of server configurations, it is up to applications to choose their specific server configuration. While tools, such as Amazon Spot Bid Advisor [2], may help users in selecting servers based on price, they expose only coarse volatility information, e.g., low, medium, or high.

## 3.1 Reducing Risk through Diversification

We use two key insights for reducing server revocation risk for a distributed batch-oriented application. The *first insight* is to choose servers which have mean time between revocations (MTTR) that significantly exceed the expected job length. For example, if a batch job has an expected length of 5 hours, then it will have a higher probability of completion if it runs on a server with MTTR of 100 hours, when compared to running it on a server with MTTR of 10 hours. Thus, choosing server configurations where the MTTR is much greater than the job length also increases the chances of a job completing without any revocations.

Each transient server configuration in a cloud platform represents a *market* with its own supply and demand conditions. If a parallel batch-oriented job chooses homogeneous servers from a single cloud market, then any revocation event will cause *all* servers to be lost simultaneously (in Amazon's EC2 spot market, if the spot price rises above bid price, then all the servers with that bid-price are revoked.). Our *second insight* for reducing the impact of concurrent revocations is to choose a heterogeneous mix of transient servers drawn from multiple markets.

Empirical analysis indicates that price fluctuations across markets are largely uncorrelated with each other (Figure 2). Thus, the revocation events in one market may not cause revocations in certain other markets, since surging demand and revocations in one market will not impact available capacity in other independent markets. As a result, use of a heterogeneous mix of transient servers drawn from independent or weakly correlated markets can mitigate the impact of revocations—since revocations now only impact a fraction of an application's servers. This enables jobs to make forward, albeit degraded, progress on the remaining transient servers.

However, constructing such a heterogeneous mix of servers from multiple markets is not trivial. It involves selecting transient servers that are "cheap" and yield high savings compared to their on-demand counterparts, yet at the same time we must minimize the risk of simultaneous revocations—if all markets fail simultaneously, there is little value in diversification. Thus we must satisfy two objectives: pick markets to minimize cost *and* minimize their failure correlation. The large number of possible markets (>2,500 spot markets on Amazon EC2), means that achieving this dual objective is intractable with ad-hoc techniques [45] and heuristics [43] that past work on multiple transient server selection has used. We describe our solution to this multiple server selection problem using portfolio theory next.

### 3.2 Server Portfolios

Intelligent server selection is key to minimizing the frequency and magnitude of disruptions seen by applications running on transient servers. To address this problem, we present *server portfolios*, a new model-driven framework to create virtual clusters composed of a mix of transient server types which offer flexible costs and availability.

Portfolios enable ExoSphere to construct a mix of cloud servers tailored to application needs. Server portfolios draw inspiration from finance [17, 37, 38]. Intuitively, a financial portfolio involves creating a suitable mix of financial investments for an investor that are drawn from an underlying mix of assets such as stocks, bonds, etc. The goal is to construct a mix that matches the investor's tolerance for risk and reward. The risk tolerance dictates whether the portfolio contains a more risky mix of high-reward assets, or a mix of lower-reward but lower-risk assets.

Similarly, server portfolios comprise a mix of transient servers that are drawn from an underlying mix of all transient server markets. Like financial assets, transient server markets exhibit different price and revocation characteristics. Some markets may have low prices but higher revocation rates, while others have higher, more stable, prices with infrequent revocations. Consequently, depending on the risk tolerance of an application, server portfolio construction involves maximizing the risk-adjusted returns by designing an appropriate mix of server markets.

ExoSphere instantiates the model-driven portfolio mechanism to create virtual clusters for applications. At startup time, applications specify their aggregate resource requirements (CPU-cores and memory) in the form of a resource vector $\mathbf{r} = [r_{\text{cpu}}, r_{\text{mem}}]$ , and their risk tolerance[1]. It then uses portfolio creation models and algorithms that are rooted in Modern Portfolio Theory [37, 38] to construct a mix of servers for the application, as discussed next.

### 3.3 Model-driven Portfolio Construction

We now present ExoSphere's portfolio model, which is based on Modern Portfolio Theory[2]from financial economics [17, 37, 38]. The goal in ExoSphere is to maximize *risk-adjusted returns* for each application, where the returns are the *cost savings* from using transient servers (over the on-demand prices), while risk is the application's tolerance to server revocation events. Formally, ExoSphere finds a suitable mix of transient servers that maximize the risk-adjusted expected return given by:

$$E[\text{Return}] - \alpha \cdot \text{Risk} \tag{1}$$

where $E[\text{Return}]$ is the difference between the cost of an on-demand server and the expected cost of the transient server. To formally define $E[\text{Return}]$, assume that the cloud platform offers servers in $n$ distinct markets. Let $D_i$ denote the on-demand price, and let $E[S_i]$ denote the mean of the transient server price. Then,

$$\text{Return}_i = 1 - \frac{E[S_i]}{D_i} \tag{2}$$

Let $\mathbf{c}$ denote the vector representing the returns for all $n$ markets, where $\mathbf{c} = [\text{Return}_1, \ldots, \text{Return}_n]$. Let $x_i$ denote the fraction of servers from market $i$ chosen in our portfolio ($0 \leq x_i \leq 1$). Then $\mathbf{x} = [x_1, \cdots x_n]$ denotes the portfolio

[1]If available, the estimated job length can be provided, and only markets with MTTR >> job-length are considered.

[1]Modern Portfolio Theory was first proposed in 1952 [37] and remains the foundational basis for much of portfolio optimization in finance even today [38].

allocation vector, and $\mathbf{x}^T$ is its transpose. The effective expected return of a portfolio is then:

$$E[\text{Return}] = \mathbf{c}\mathbf{x}^T \tag{3}$$

The parameter $\alpha$ (in Equation 1) denotes the *risk-averseness* of the application or user. A low value of $\alpha$ indicates that the application places lower emphasis on avoiding server revocation risk. Conversely, a high value of $\alpha$ indicates that an application is highly risk-averse, and is willing to incur an extra cost for this. We also use the term risk tolerance to mean the inverse of risk-averseness.

To capture risk, we draw an analogy with financial portfolio selection, where investments are chosen such that their prices are not correlated. The rationale is that if one asset (say, a particular stock) sees a decline in price, then the other assets (e.g., a bond) are unlikely to see a concurrent decline. This way, we avoid large declines in the overall portfolio value.

In our case, we wish to select server markets with independent revocation events—thus if there is a revocation in one market, others will not see a concurrent revocation. This reduces the total number of allocated servers that are revoked. To do so, we define a covariance matrix $\mathbf{V}$ that captures pairwise correlations between all pairs of markets. $V_{ij}$ is the correlation between markets $i, j$, and captures their simultaneous revocations. Higher values indicate that the two markets are highly correlated in their revocations, and the chances of closely spaced revocations are greater. We use this formulation to define the revocation risk of a portfolio as:

$$\text{Risk} = \mathbf{x}\mathbf{V}\mathbf{x}^T \tag{4}$$

Our portfolio construction problem can then be formulated as the following optimization problem:

$$\begin{aligned} \text{Maximize:} \quad & \mathbf{c}\mathbf{x}^T - \alpha\mathbf{x}\mathbf{V}\mathbf{x}^T \\ \text{Subject to:} \quad & \sum_{1}^{n} x_i = 1 \\ & \mathbf{x} \geq 0 \end{aligned} \tag{5}$$

We can solve Equation 5 for a wide range of risk-aversion parameters ($\alpha$) to compute the lowest-cost portfolios for any given risk. The expected returns and revocation risks of these portfolios are shown in Figure 3, which shows the expected cost savings for a range of revocation risks. As the revocation risk is reduced, so is the cost savings. We also see from Figure 3 that expanding the candidate-set from r3 servers in the US-east-1 region to *all* the servers in the US-east-1 region results in a 1% increase in savings, and a 20-50% reduction in revocation risk. This occurs because a larger set of candidate markets both allows more freedom in choosing markets, and increases the number of markets with low correlations.

The effectiveness of the risk-averseness parameter can also be seen in Figure 4, which shows the distribution of servers in portfolios with different risk-averseness parameters. We can see that the portfolios become more diversified as the risk-averseness increases.

**Constructing the covariance matrix.** The covariance matrix $\mathbf{V}$ captures the pairwise correlation between markets. Our formulation allows multiple types of correlation to be used. The different correlation functions (and their corresponding $\mathbf{V}$ matrices) allows ExoSphere to adjust the portfolios to the users' perceptions of risk.

The first and most basic form of correlation is simply the correlation between the spot prices. In the case of Amazon EC2, we can use price histories of spot servers, which are publicly available, to compute the mean returns and the covariance matrix. That is, we compute the pairwise covariances by using spot prices to capture revocation events and using the standard covariance formulation. Let $X_t, Y_t$ denote the spot price of markets $X, Y$ respectively at time $t$. Then the standard definition of covariance applies:

$$V_{XY}^{price} = \frac{1}{T}\sum_{t=1}^{T}(X(t) - E[X])(Y(t) - E[Y]) \tag{6}$$

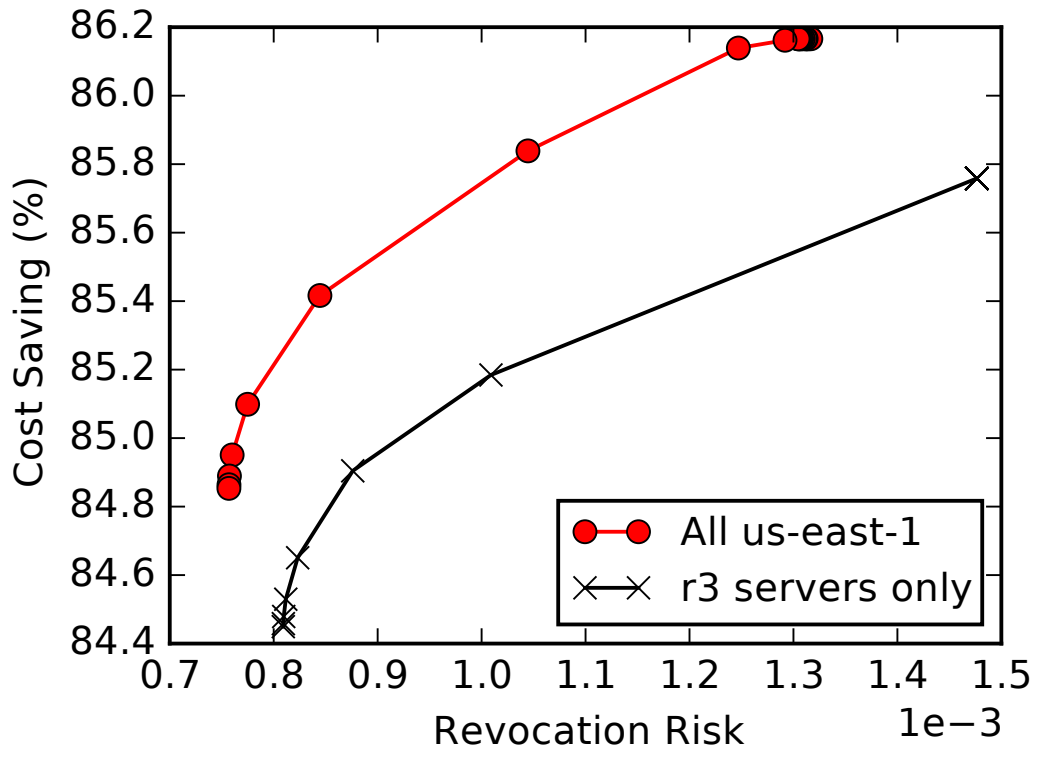

Fig. 3. Cost savings and revocation risks of portfolios with different risk-averseness. Choosing a portfolio from a larger collection of servers (all US-east-1 vs. only r3-type servers) results in higher returns at lower risk.

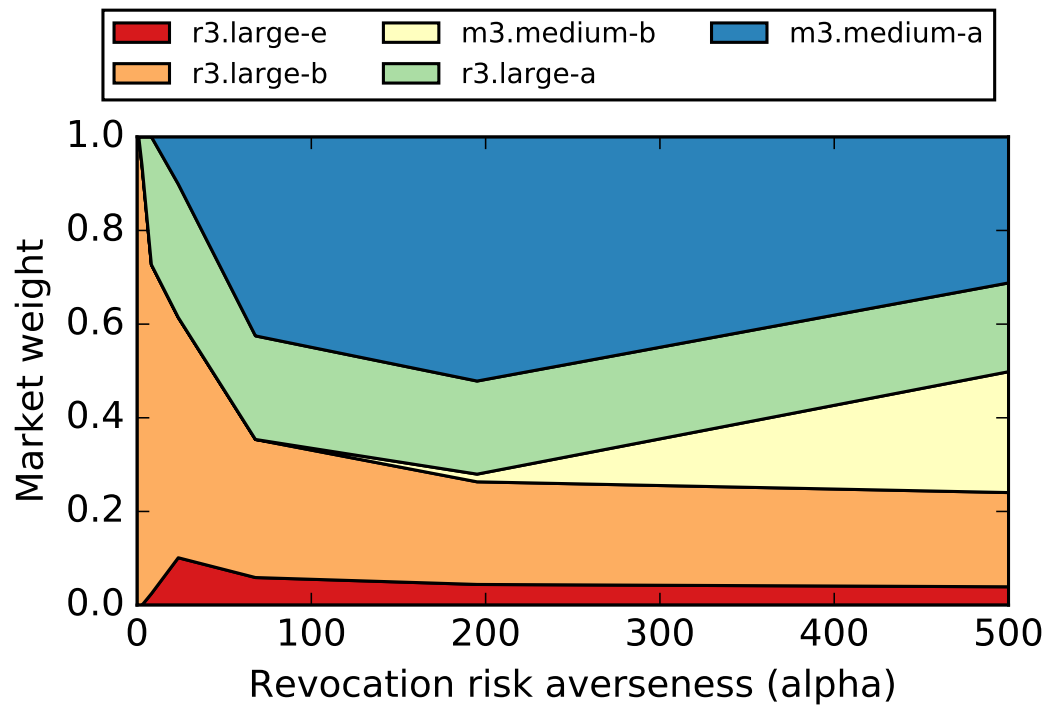

Fig. 4. The effect of risk-averseness in portfolio diversity. A single market (r3-large-b) dominates the portfolio when $\alpha = 0$, but the portfolio's diversity increases with increasing risk-averseness.

This captures the correlation between the prices in different markets, and is useful metric for price sensitive users, since they may not want prices of all markets to increase simultaneously.

In transient server environments, simultaneous market revocations can lead to disruption of application availability or performance. To capture simultaneous revocation risk between two markets, we use the likelihood of simultaneous revocation. We again use the spot price traces to find simultaneous revocations between markets—we say that two markets have a simultaneous revocation if servers in those markets get revoked within a small time window (5 minutes). This allows us to define the entries in the covariance matrix using the probability of simultaneous revocations.

$$V_{XY}^{revoc} = \text{Probability of simultaneous revocation of X,Y} \tag{7}$$

Lastly, we also provide a hybrid risk formulation that captures both simultaneous revocations and changes in prices. We first transform the spot prices to capture revocation events, and then compute the covariance of these transformed prices. Since a spot server is revoked if its price increases above the bid price, we capture the revocation and unavailability by setting the price to the maximum spot price. For a given market, if we are given a trace of the spot prices $S$ and the bid price $B$, we define the transformed prices as:

$$\begin{aligned} S'(t) &= S(t) && \text{if } S(t) < B \\ &= \text{Maximum spot price} = 10 * \text{On-demand price} && \text{otherwise} \end{aligned}$$

This ensures that we impose a very high uniform penalty when there are revocations. Because we set the prices to the maximum spot price during a revocation, this results in a high correlation if two markets fail at near the same time. The final step is to compute the covariances between pairwise markets (after applying the above price transformation) by using the standard covariance formulation : $V_{ij}^{hybrid} = \frac{1}{T}\sum_{t=1}^{T}(S_i'(t) - E[S_i'])(S_j'(t) - E[S_j'])$.

Our portfolio model (Equation 5) formulation closely mirrors portfolio construction found in Modern Portfolio Theory. Infact, it is a quadratic convex optimization problem.

First, we note that the covariance matrix $\mathbf{V}$ is positive semidefinite. $\mathbf{V}$ is a matrix of covariances that are always non-negative. To establish semidefiniteness, it is enough to show that for any vector $\mathbf{a}$, $\mathbf{a^T V a} \geq 0$. By using the definition of co-variance, we get:

$$\begin{aligned}\mathbf{a^T V a} &= \mathbf{a^T} E[(x-\mu)(x-\mu)]\mathbf{a} \\ &= E[(\mathbf{a^T}\cdot(x-\mu))((x-\mu)\cdot\mathbf{a})] = E[((x-\mu)\mathbf{a})^2] \geq 0\end{aligned}$$

Since the covariance matrix is positive semidefinite, $\mathbf{x^T V x}$ is strictly convex, and thus the problem formulation in Equation 5 is a quadratic convex optimization problem [16, 38]. The formulation can be solved by an off-the-shelf convex solver, such as `cvxopt` [9]. This allows us to exactly solve the portfolio modeling problem and get portfolios that maximize the risk adjusted returns, without having to rely on heuristics or approximation. For Amazon EC2 spot instance portfolios, we use the publicly available time series of spot prices for each spot market. We can then compute the average spot price for each market and can get the returns vector **c**, as well as the covariance matrix **V**.

## 3.4 Server Allocation using Portfolios

ExoSphere considers the risk-averseness requirements of the application along with the computing resource requirements. Based on these requirements, ExoSphere first constructs a portfolio of resources on cloud servers, and then allocates the resources to the applications in the form of containers on these servers.

Applications submit CPU and memory resource requests in the form a resource-vector $\mathbf{r} = (r_{\text{cpu}}, r_{\text{mem}})$, and their placement constraints. The placement constraints comprise primarily of the risk-averseness factor $\alpha \in [0, \text{inf})$, and any server preferences they might have (gpu-enabled servers only, no small servers, etc).

We then construct portfolios based on these requirements, which gives us the weights for each market in the form of a weight-vector **x**. These weights represent the fraction of resources that must be allocated in a market. For each market $i$, we compute the CPU and memory resources that must be allocated in that market by multiplying the portfolio-weight of that market ($x_i$) by the resource-vector (**r**). ExoSphere then determines the actual number of servers to allocate in market $n_i$ based on the CPU and memory capacities of the servers in that market ($\text{CPU}_i, \text{MEM}_i$) as follows:

$$n_i = \max\left\{\frac{x_i r_{\text{cpu}}}{\text{CPU}_i}, \frac{x_i r_{\text{mem}}}{\text{MEM}_i}\right\} \tag{8}$$

We take the maximum of the servers required to satisfy both the CPU and memory requirements so that the application's resource allocation meets or exceeds the requirements in all resource dimensions. This approach can be extended to other resource types (disk/network bandwidth, etc.). Upon deciding the number of servers that an application needs in each market, ExoSphere then requests new servers (with bid price set as the on-demand price) from the cloud operator. ExoSphere also allows applications to dynamically adjust their resource requirements, which is useful for auto-scaling. Applications can adjust their CPU and memory requirements (**r**) at any time, and ExoSphere adds or removes servers from each market.

## 3.5 Statistical Multiplexing of Servers

In the above described server allocation policy, it may be possible for an application's resource requirements to be smaller than the resources offered by the server portfolio. This can occur because of two reasons. The first reason is that ExoSphere maximizes the (risk adjusted) cost-savings relative to the on-demand price, which may require selection of larger servers. Such price inversions are common in EC2 spot markets, and can occur if smaller transient servers have a larger demand compared to their larger counterparts. The second reason for surplus resources in a portfolio is that ExoSphere's allocation ensures that sufficient servers are available to meet the demands across all resource types i.e., both CPU and memory. For example, an application requesting 2 CPUs and 10 GB memory may be allocated a portfolio of 2 `m3.large` servers each having 2 CPUs and 7.5 GB memory, resulting in 2 free CPUs and 5 GB of free memory across both the servers.

ExoSphere reduces the surplus unused resources in a portfolio by relying on statistical multiplexing. The key idea is that transient servers can be multiplexed across multiple portfolios. This allows multiple applications to share the servers in their virtual clusters such that the free and unused resources of a server can be used by other applications. In addition to increasing server utilization, this also reduces costs, since the cost of transient servers is also proportionally shared between the applications sharing a server.

ExoSphere's statistical multiplexing, also referred to as the shared-cluster policy, works as follows. We use the portfolio modeling and creation process described earlier. This gives us the portfolio weights vector **x**, indicating the weights of each market in the portfolio. The application's actual resource requirements (**r**) are first met by trying to use as many surplus resources as possible across all the servers in a given market. That is, for each market in the application's portfolio, we first find surplus resources on existing servers in that market, and then request the cloud servers required to meet the unmet resource demand in that market instead of all $n_i$ servers (Equation 8). Finding surplus resources involves finding servers such that their allocated-resource vector is less than the available resources. ExoSphere uses the "best-fit" policy: it sorts the servers in each market in descending order of their free resource availability, and then proceeds to allocate resources (as containers) from these servers until either all free resources in the market are allocated or if the application's resource requirements in that market are satisfied.

Finally, we note that this multiplexing of servers is only effective if there exist multiple applications to exploit the free resources, and if there is a steady stream of applications leaving and entering a system. We evaluate the cost effectiveness of this multiplexing scheme in Section 6.3. In the next section, we describe how applications can use the API provided by ExoSphere to design and implement their own transiency-specific policies.

## 4 APPLICATION INTEGRATION

In addition to supporting the portfolio abstraction, ExoSphere provides a number of key mechanisms to support the execution of batch-oriented applications on transient servers. ExoSphere's design is based on the Exokernel philosophy, where it provides a small set of mechanisms to make an application transiency-aware, and leaves the design of transiency-specific policies to the application.

Unlike much of prior work on running applications on transient servers, ExoSphere gives applications the ability to define their own policies for handling revocations. This allows applications to define policies to suit their fault tolerance requirements, and also allows more efficient fault tolerance. For example, using application-level fault tolerance such as application-level checkpointing [43] may significantly reduce the overhead of checkpointing compared to application-agnostic system-level checkpointing.

ExoSphere uses a two-level architecture (Figure 5), where ExoSphere provides the portfolio abstraction and transiency-specific "up-calls" to the applications, which may use them to implement their own policies. Associated with each application is a job-manager, which communicates with ExoSphere to implement these policies.

Given any vanilla batch-oriented application, converting it into a transiency-aware variant of that application involves defining three policies: a (i) *portfolio policy*, which specifies its resource needs and risk tolerance, (ii) *fault-tolerance policy*, which specifies whether and how the application state is saved to deal with potential server revocation, and (iii) *recovery policy*, which specifies the policy to replenish servers upon a revocation event and to resume the application after recovering saved state.

The portfolio policy is implemented using ExoSphere's portfolio abstraction described in the previous section. To implement a broad range of fault-tolerance and recovery policies, ExoSphere supports three key mechanisms via the up-call API described in Figure 6:

**Exposing the Portfolio MTTRs**: Since cloud platforms only expose transient server prices but not revocation statistics, ExoSphere provides MTTR information immediately after portfolio creation and periodically (every 5 minutes) via the `portfolioMTTR` upcall. ExoSphere provides the mean MTTR of an application's portfolio, as well as the specific MTTRs of the individual transient servers within the portfolio. An application can use this knowledge of how frequently a portfolio server is likely to be revoked to tune how frequently to save its state.

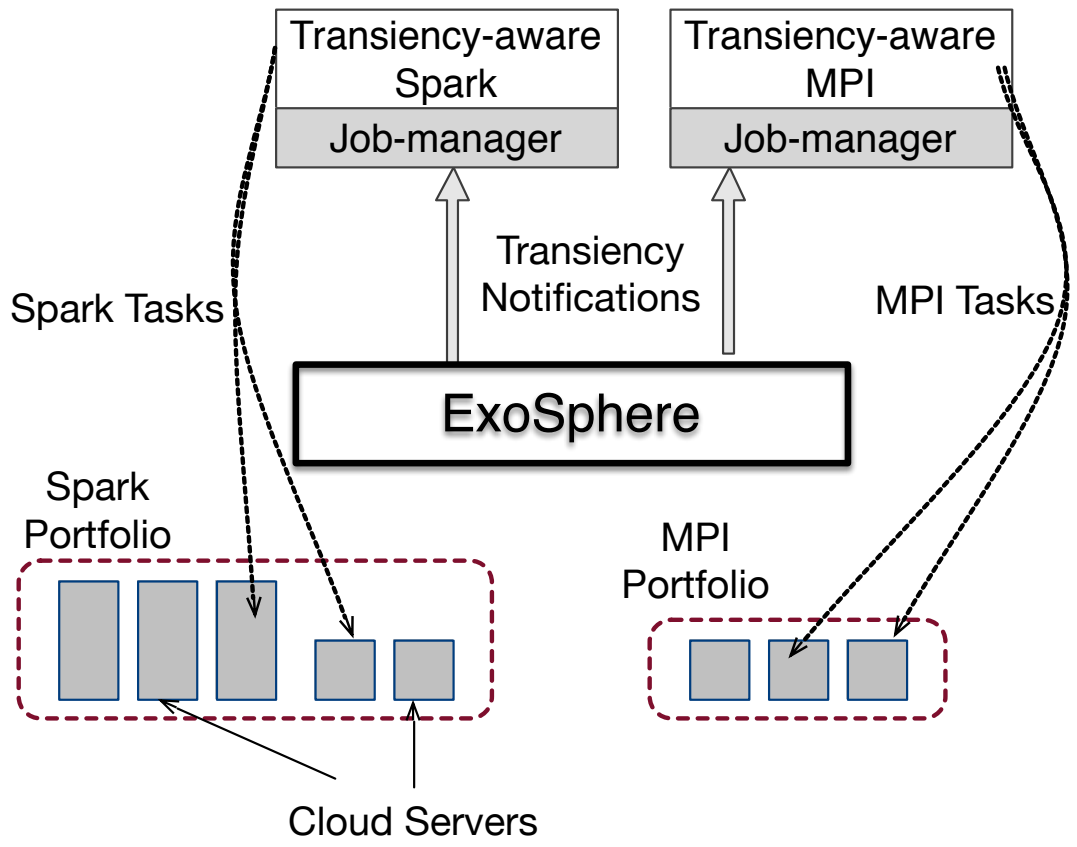

Fig. 5. ExoSphere's design architecture. The job managers for each application implement the resource allocation requests and the fault tolerance policies.

| Down-Calls |
|---|
| `portfolio = reqResources(cpu, mem, alpha, job-len)` |

| Up-Calls |
|---|
| `mttrList = portfolioMTTR(portfolio)` |
| `Notify(`**`event`**`, serverList)` <br> **`event`** `= {hardRevocation, softRevocation, priceThreshold}` |

Fig. 6. ExoSphere API. Applications allocate portfolios by using down-calls, and receive transiency-specific notifications using the up-calls provided by ExoSphere.

**Hard revocation signals**: ExoSphere tracks a cloud platform's termination warning for one or more servers and signals the application about imminent revocation of these servers via the `Notify(hardRevocation)` upcall.
**Soft signals**: Soft signals are provided to signal specific conditions to the application. ExoSphere currently supports two types of soft-signals: (i) price threshold signals and (ii) soft revocation. Price threshold signals are used by price sensitive applications to track when the price exceeds a specified threshold, by using the `Notify(priceThreshold)` upcall, which it can use to relinquish servers and restart computation later to avoid going over budget. Soft revocation signals are upcalls from ExoSphere when it detects early signs of revocation—they serve as an early warning (but not a guarantee) that revocation may occur in the near future (e.g., when the signature of a price spike is detected). The soft-revocation notification provides more time for applications to take action (e.g. checkpoint, migrate, etc).

Next, we describe how these mechanisms can be used to create transiency-aware versions of three common batch-oriented applications with modest effort.

## 4.1 Data-parallel Application: Spark

The low-cost of transient servers makes them very appealing for running data-parallel data-intensive frameworks like Hadoop, Spark, Naiad, etc. Such frameworks run two broad classes of jobs. Traditionally, they run data intensive batch jobs that perform computation over large amounts of data in parallel. These frameworks also support batch-interactive [43] jobs such as SQL queries, interactive machine-learning, or streaming analytics, which have lower-latency requirements.

We use Spark [60] as a representative data-processing framework to build a transiency-aware application using ExoSphere. Spark is a popular data-parallel framework and supports both batch and batch-interactive computation. Spark performs data transformations on in-memory distributed datasets called Resilient Distributed Datasets (RDDs [60]). The loss of servers leads to the loss of the in-memory RDD partitions, which can lead to recursive recomputation. While batch workloads can tolerate the delay due to recomputation, such recomputation significantly increases the latency for batch-interactive workloads, such as SQL queries or REPL-based environments.
**Portfolio Policy.** The portfolio policy for a Spark cluster depends on the Spark workload characteristics. Purely batch workloads are more disruption tolerant and may choose to optimize for lower cloud costs. Thus, when instantiating a Spark cluster for a batch workload, a low risk-averseness portfolio (low $\alpha$) can be requested. Doing so will skew the portfolio towards lower costs. In contrast, batch-interactive and streaming workloads are highly risk-averse, and thus request highly diversified portfolios with a high $\alpha$ to reduce the performance impact of revocations (but at potentially higher cost).

**Fault-tolerance Policy.** Spark includes a built-in RDD checkpointing mechanism, which serializes RDDs to stable storage. However, Spark leaves it to the application to decide which RDD to checkpoint. A checkpoint operation imposes significant overhead, since it causes a substantial amount of in-memory data to be written to disk.

Designing a fault-tolerance policy for our transiency-aware version of Spark is straightforward using this checkpoint operation—we periodically checkpoint recent RDDs. Due to the overhead imposed by checkpointing, the checkpoint interval must be carefully chosen. Since ExoSphere exposes the MTTR of the portfolio, we can use it to set the checkpointing interval to $\tau = \sqrt{2 \cdot \delta \cdot \text{MTTR}}$, where $\delta$ is the time it takes to write a checkpoint to disk, and the MTTR is Mean Time To Revocation of the portfolio. This expression follows directly from a classic result in fault-tolerance [21] and has been used in other Spark-based systems such as Flint [43]. To implement this policy, we modify the Spark job-manager to periodically checkpoint RDDs, and use saved checkpoints when resuming after a revocation. The pseudo-code for the Spark periodic checkpointing is below:

```
while(true):
  mttr = portfolioMTTR(portfolio).get()
  tau = math.sqrt(2*mttr*delta)
  sleep(tau)
  for rdd in job.rdds.SortBy(``age'')[0] :
      rdd.checkpoint()
```

**Recovery Policy.** The recovery policy comprises of two parts: how to recover the application upon a revocation event, and how to resize the cluster to handle lost servers. Upon receiving a hard revocation signal from ExoSphere, the job-manager in our transiency-aware Spark triggers recomputation from the last saved RDD checkpoint. The decision on whether to replenish lost servers depends on the job progress and workload characteristics. Due to Spark's in-built fault-tolerance mechanisms, jobs are able to continue execution on remaining servers. However, continuing in this degraded mode increases job completion times (even when resuming from a saved checkpoint), due to the potential of spilling RDDs to disk, or reduction in the size of the RDD cache.

For pure batch jobs, we can use job progress (by comparing against estimated job-length), and intelligently decide whether to replenish (e.g., replenish if job-progress $< 70\%$). For batch-interactive or streaming workloads, an immediate replenishment policy is always preferred due to the latency requirements.

**Comparison with other Spark-based systems.** Flint [43] and TR-Spark [57] are two recently proposed transiency-aware versions of Spark. Both systems use an application-level fault-tolerance and require significant complex modifications in Spark to embed new mechanisms and policies. Our version uses ExoSphere abstractions and mechanisms to implement similar policies. We model our version on Flint's design. However, while Flint requires 3000 lines of code changes [43] to Spark, ExoSphere requires adding only 400 lines, and benefits from separating issues such as portfolio construction out of the application. ExoSphere also allows a richer server selection policy, since portfolios can be tailored to the workload's risk tolerance.

TR-Spark [57] is another attempt to make Spark transiency-friendly, and changes task-scheduling in Spark to avoid scheduling jobs to nodes that face imminent revocation. These changes can also be supported by ExoSphere, since TR-Spark also uses MTTR information. Mostly, ExoSphere's Spark benefits from separation of concerns and requires less changes to the application (Spark) in order to run on transient servers.

## 4.2 Parallel HPC Application: MPI

Message Passing Interface (MPI) is the predominant framework for scientific and high-performance computing. MPI jobs tend to be parallel compute-intensive tasks and their large degree of parallelism can benefit from running on low-cost transient cloud servers [36]. However, unlike Spark, MPI's message-passing model is highly intolerant to revocations. In particular, revocation of a single server can cause the entire MPI job to fail.

**Portfolio policy.** Since even a single server revocation requires the entire job to be restarted (from the beginning or from a checkpoint), a policy that attempts to limit failures to a fraction of the servers is not adequate—any revocation, whether it is one server or all servers, has the same impact. Thus, stability is more important than server diversity, i.e., choosing servers with MTTR $>>$ the job length, which reduces the probability of revocation, is more important than portfolio diversity. Thus, MPI's job-manager requests portfolios by specifying the expected job-length, and specifies a *low* risk-averseness parameter to ensure selecting high-MTTR servers.

**Fault-tolerance policy.** Many MPI platforms, such as OpenMPI [8], support checkpointing. In such cases, the MPI job can periodically checkpoint its state similar to ExoSphere's Spark. If checkpointing is not supported or is undesirable, then no fault-tolerance policy is necessary and the job is simply restarted from the beginning.

**Recovery policy.** Due to the inability of MPI jobs to continue computation after partial failures, the immediate replenishment policy must be used to restore the cluster to its original size upon a failure of one or more servers. Once replenished, the job is restarted from the most recent checkpoint or the beginning. The pseudo-code of the revocation-handling policy for MPI is shown below:

```
def Notify(hardRevocation, servers):
  Kill_MPI_Job()
  portfolio = reqResources(cpu, mem, alpha=0)
  Start_MPI_Job(portfolio)
```

The ExoSphere MPI version required a modest effort of 50 lines of code for the portfolio and recovery policy.

## 4.3 Delay Tolerant Application: BOINC

Volunteer computing frameworks such as BOINC [12] are an example of "embarrassingly parallel" workloads that are delay-tolerant and do not have strict deadlines.

**Portfolio policy.** Since reducing cost is more important than mitigating failures, a low-to-moderate risk-averseness parameter ($\alpha$) can be specified when constructing a portfolio for BOINC. For highly price-sensitive workloads, a low value may be used, but it risks losing a large fraction (or all) servers. Use of a moderate value provides some diversification, which allows progress to be made when part of the portfolio is revoked.

**Fault-tolerance policy.** Typically no fault-tolerance mechanisms are needed, since if a server is lost in a volunteer-computing scenario, the task is restarted. In some cases with long-running tasks, a lazy-checkpointing policy can be used, which checkpoints the task after receiving a soft or hard revocation warning. Soft warnings increase the chances of completing the checkpoint, since a lazy-checkpoint may not complete within the hard-warning duration (2 minutes on EC2, 30 seconds on GCP).

Due to its price sensitive nature, BOINC can use soft signals to set a price threshold and upon receiving a notification of rising prices, can voluntarily relinquish servers and wait for the price to reduce to maintain a budget.

**Recovery policy.** The price sensitive nature implies that immediate replenishment of lost servers is not strictly necessary. The BOINC job-manager can monitor the price of portfolios offered by ExoSphere to wait until prices drop. Tasks that were affected due to server revocations are simply queued on other remaining nodes and are restarted (from the beginning or from the last checkpoint).

The transiency-aware BOINC required about 200 lines of additional code—most of which pertain to the implementation of lazy checkpointing and recovery.

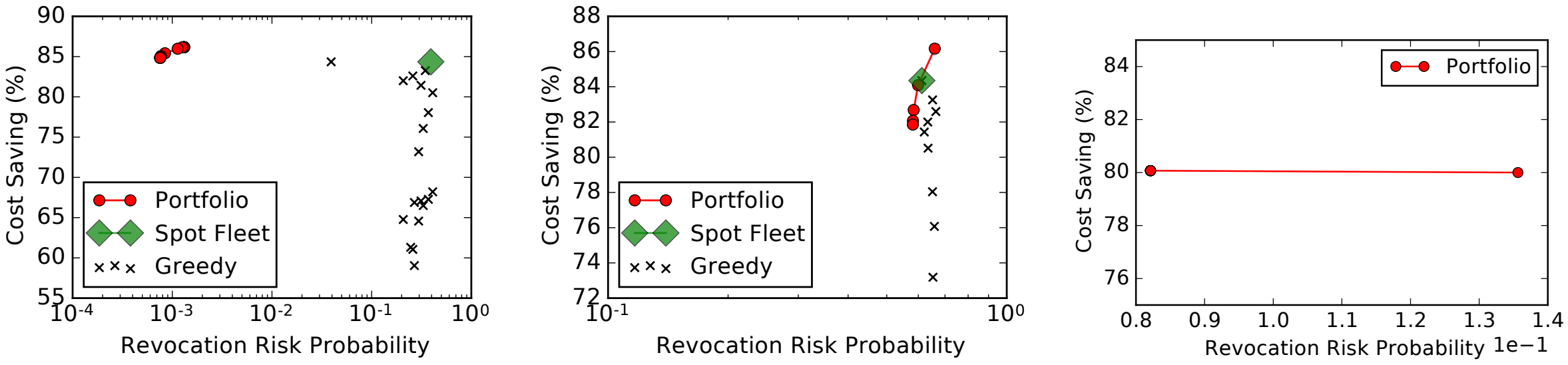

(a) Today's EC2

(b) Synthetic highly-correlated EC2 markets

(c) Google preemptible instances

Fig. 7. *Portfolios for various market scenarios*

## 5 EXOSPHERE IMPLEMENTATION

While ExoSphere's design and its portfolio mechanism are general, we implement ExoSphere using Mesos [32]. The choice of an existing cluster manager for implementing ExoSphere is motivated by two factors. First, Mesos employs an Exokernel [27] like philosophy of letting higher-level applications implement their own specific resource allocation policies. Thus, ExoSphere's abstraction and interfaces are a natural fit into the architecture of such cluster managers. Second, enhancing a popular cluster manager such as Mesos to support transient cloud servers yields a transiency-aware cluster-manager that can find broad use and adoption in today's cloud platforms.

ExoSphere is built using Mesos v0.27 and cloud native APIs. Our prototype has two key components, the ExoSphere master and the application job-managers. The ExoSphere master is implemented in 5000 lines of C++ code by extending the Mesos master. The master implements two key components: portfolio-based resource allocation and the application-facing API shown in Figure 6. Our prototype currently supports EC2 spot instances, which offer a rich collection of servers and publicly available price history. We also have proof-of-concept support for Google's preemptible instances (where inferring availability information is more challenging).

ExoSphere requires applications to implement a job-manager using ExoSphere's API to design their own portfolio-creation, fault-tolerance, and recovery policies. Requiring applications to implement their own job-manager is increasingly common in modern cluster managers—our job-manager is equivalent to Application-Schedulers in Mesos and Application-Masters in Yarn. The master communicates with the job-managers using the existing Mesos RPC and HTTP APIs. This allows existing application-schedulers (written for Mesos) to be used and augmented with the transiency-specific functions provided by ExoSphere. The only requirement for running existing Mesos applications on ExoSphere is that they handle the revocation hard-warning notification, which can be implemented in either C++, Java, or Python.

Applications (via their job-managers) make requests for resources and their portfolio requirements via the existing Mesos `requestResources` RPC, which the ExoSphere master intercepts and handles. The resource allocation involves portfolio construction, server-packing, and creating and keeping track of the cloud servers.

**Portfolio-based allocation.** In order to construct portfolios, we use historical price traces. Amazon publishes the past three months of spot price traces (available using the EC2 `describe-spot-price-history` API). We periodically collect the price history for all markets, and compute the mean spot price, as well as the covariance matrices for various risk functions. Once these have been computed, ExoSphere solves the quadratic convex optimization problem using the Python `cvxopt` solver, which takes under 1 second for the 250 us-east-1 markets and under 25 seconds for the 2500 global markets. The portfolios for various risk-averseness factors are precomputed and cached, and this reduces the computational overhead of portfolio construction even further. In the absence of any server-type or job-length constraints, portfolio construction usually only involves a simple look-up/search in the portfolio cache.

ExoSphere does explicit, fixed resource allocation, and does not use Mesos's Dominant Resource Fairness allocator. Once the application terminates or voluntarily relinquishes its resources, its servers are placed on a free-list of servers for a short duration ($2\times$ allocation latency), instead of immediately terminating them. Similar to anticipatory scheduling, holding on to recently relinquished servers in the free-list speeds up the allocation of servers for future applications, since launching transient servers takes a few (~5) minutes.

New cloud servers are requested using the standard EC2 APIs, and are started with either the application provided disk-image (containing the required application dependencies), or a default image (AMI) which has a few common applications installed. We assume that most applications will use S3 or EBS for storing data, since the content of local disks is lost upon server revocation. The resources on cloud servers are offered to the applications using the Mesos abstraction of resource offers.

ExoSphere's portfolio-based policy may over-allocate resources, which can lead to idle resources on some servers. For example, an application requesting 2 CPUs is allocated a cloud server with 4 CPUs results in 2 surplus CPUs. To increase cluster utilization and reduce costs, ExoSphere also implements a server packing policy as an optimization, which first tries to meet resource demands of the application (in each market) from the idle resources on the servers in that market. For this, we use a simple first-fit approach to allocate resources. Note that applications run inside containers (e.g., Mesos executors), which provide security and performance isolation. Nevertheless, applications which do not wish to face the potential interference because of other co-resident applications can still request private cloud servers not shared with other applications.

**ExoSphere Upcalls.** The ExoSphere master also interacts with the servers and the cloud provider in order to issue transiency-specific notifications. Revocation hard-warnings are first detected by the servers, which then inform the master, which relays them to the applications via the Mesos `inverseOffers` API, which includes a list of affected servers/containers and the remaining time until termination. Soft revocation warnings are provided by monitoring the state of each server, and notifying the application if it reaches the `marked-for-termination` state. Additionally, the master can also bid much higher than the on-demand price and monitor for price increases to increase the soft-warning duration. Price notifications are used by applications to know if the price of their portfolio has increased above a threshold. The ExoSphere master uses the `describe-spot-price-history` EC2 API to continuously monitor prices of all the active markets and delivers the notification if the price crosses the threshold.

## 6 EXPERIMENTAL EVALUATION

Our experimental evaluation focuses on answering two key questions: i) What is the effectiveness of the portfolio abstraction in reducing cost and revocation risk? and ii) What is the impact of different policies for handling revocations and with different risk tolerances? We evaluate ExoSphere on EC2, and also show results for Google Cloud Platform (GCP). We use spot price traces over a six month period (Apr-Sept 2015) for evaluating portfolios, and restrict ourselves to a single region (us-east-1), since many applications have geographic locality constraints that prevent using servers from multiple regions. We evaluate the performance of transiency-aware variants of Spark, MPI, and BOINC on ExoSphere.

**Spark.** We use the Spark version modified to work with ExoSphere, and use Amazon S3 for storing input/output data and RDD checkpoints. We use a combination of batch and low-latency workloads for Spark. We use KMeans, an iterative machine-learning algorithm with $\sim$16GB of input data as a batch workload. For the low-latency, interactive scenario, we use TPC-H database queries served by a Spark application. Spark supports SQL queries by translating them into equivalent RDD operations, with each query akin to a short running job.

**MPI.** We use MPICH [7] v2.7, which supports Mesos. We use MPI as an example of a "rigid", transiency-agnostic application, which responds to revocations by requesting new servers and restarting its job. We use the NAS parallel benchmark [13] as an MPI workload.

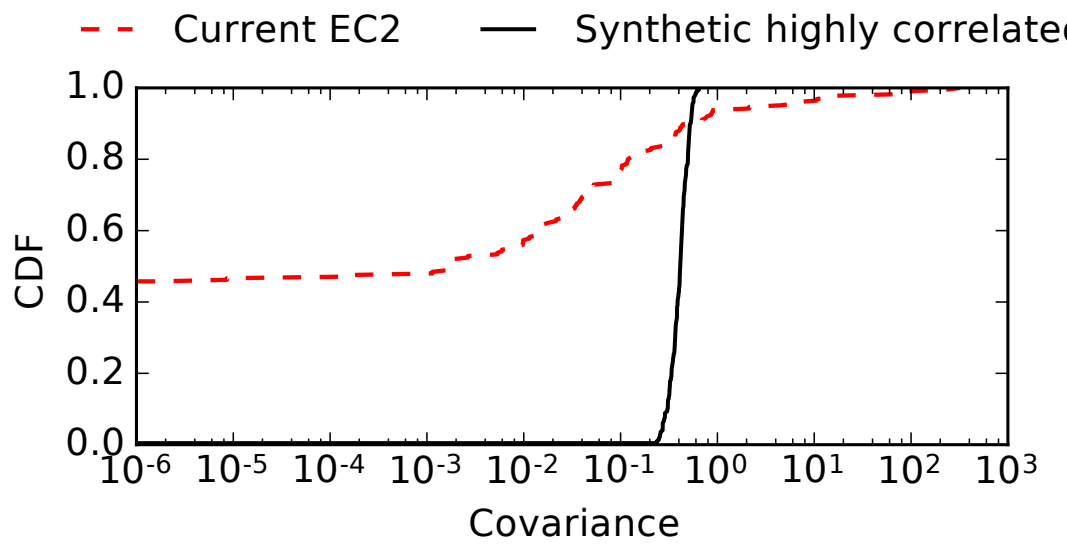

Fig. 8. CDFs of covariances.

| | Checkpointing | Restart |
|---|---|---|
| Eager-Replenish | +7% | +32% |
| No Replenishment | +11% | +54% |

Table 1. Increase in running time of Spark (KMeans workload) due to a revocation event.

**BOINC.** We use BOINC as a delay-tolerant bag-of-tasks distributed application. We configure BOINC to run fixed-length CPU-intensive embarrassingly parallel tasks.

### 6.1 Portfolio Construction

We first examine the risk-return tradeoff in portfolio construction, and compare it to other server selection approaches. A baseline approach to server selection in multiple markets is the greedy strategy which picks server-types offering the lowest average spot price. To select multiple ($k$) markets, the greedy approach picks the top-k cheapest server-types for use by the application.

Figure 7a shows the expected cost savings (compared to using on-demand servers) and the expected revocation risk for both the greedy approach and our portfolio abstraction. Since there is no explicit way to specify risk-averseness in the greedy approach, we use a cruder "number of markets" as the diversification criterion, and divide servers among all markets equally. Each point in Figure 7a for the greedy scenario refers to a different number of markets picked. The *best* greedy portfolio approaches the cost-savings offered by the portfolio approach, but yields a much greater revocation risk (by 50×). Compared to portfolios, other greedily constructed selections offer upto 40% less saving at 100× more revocation risk.

We also compare against Amazon Spot Fleet [3], which provides a risk-agnostic mix of servers. Given a set of spot server types, Spot Fleets are constructed by either of two policies: the lowest-cost policy picks the server with the lowest spot price; and the diversified policy equally distributes servers among all chosen markets. The diversified policy is thus equivalent to our greedy policy when all servers in a region are considered. The result for the lowest-cost Spot Fleets is also shown in Figure 7a. The Spot Fleet has similar cost to our portfolio approach, but has almost 100× higher revocation risk for the single-market lowest-cost policy.

In practice, a system involving greedy server selection would have to iterate over all market sizes and pick the best performing greedy selection [43], and would have no explicit way to control the revocation risk of the selection.

So far we have seen the expected behavior of portfolios in EC2's spot market. We now explore their behavior in other scenarios. Figure 8 shows the CDF of the covariances of the EC2 spot markets, which shows that there is a large number of extremely low-correlation markets with a few highly-correlated markets. If usage of spot instances and the portfolio-approach were to increase, it is possible that the increased demand for the servers in the "uncorrelated" markets would increase prices and revocations. To evaluate this scenario, we construct a synthetic highly correlated covariance matrix which has no uncorrelated markets (shown in Figure 8). For our experiment, we use these synthetic covariances and EC2 prices. Figure 7b shows that in highly correlated markets, the portfolio approach outperforms the greedy approach by 10% in both savings and risks, and is comparable to Spot Fleets. This is because the correlated markets reduce the possibility of exploiting market independence for portfolio diversification. We emphasize that this is a worst-case scenario. Today's markets provide ample opportunity to diversify across independent markets.

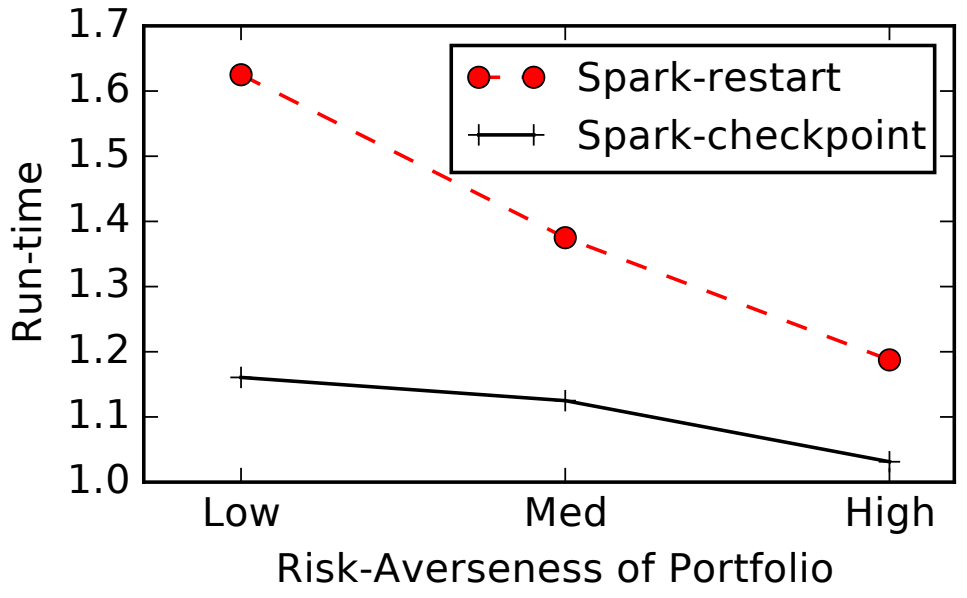

Fig. 9. Spark job run-time (normalized to on-demand servers) when one market fails.

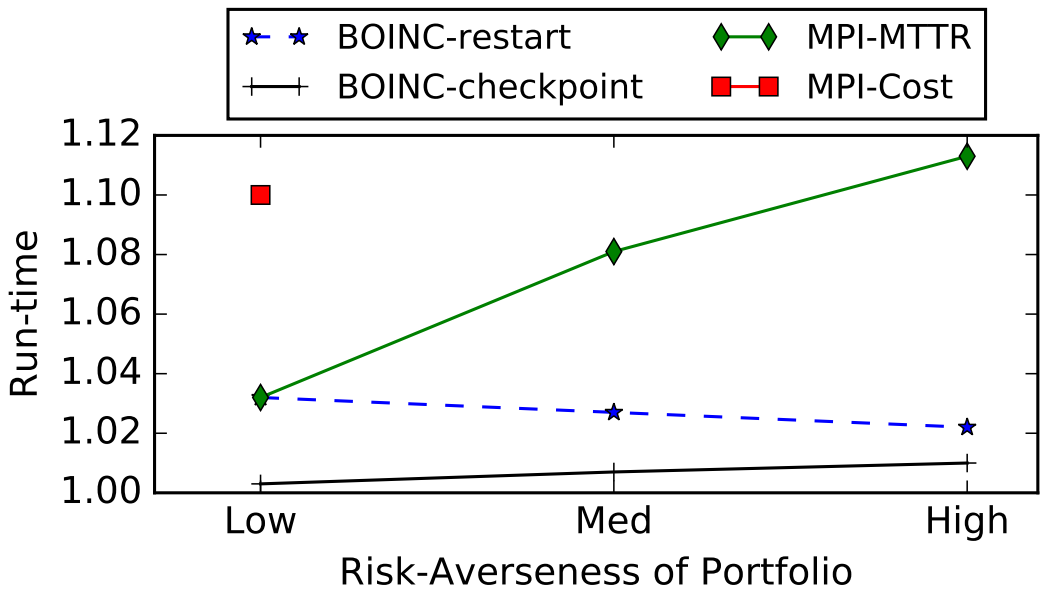

Fig. 10. BOINC and MPI and risk averseness.

**Google preemptible instances.** Unlike spot instances, Google's preemptible instances have fixed discounts. The lack of variable pricing means that we cannot use it to estimate the correlations between different server types. Given the lack of any availability data, we assume that ExoSphere would require active probing [39] to infer availability. For our experiments, we use the covariance matrix from the EC2 markets. The resulting cost savings and revocation risks are shown in Figure 7c. Because the preemptible instances have discounts in a very narrow range (80-80.2%), the portfolios also have nearly constant cost savings. Importantly, increasing diversification reduces the revocation risks by 75%.

**Result:** *Portfolio construction outperforms greedy selection in uncorrelated markets, and its diversified portfolios reduce revocation risk even when markets are correlated.*

## 6.2 Application Performance

We now examine application performance when running on portfolios with different risks and transiency-specific policies. For ease of exposition, we group portfolio risk-averseness into three types: low, medium, and high. Low risk-averseness typically corresponds to single-market mode of operation, while high risk-averseness corresponds to using as many markets as possible. Note that in practice, the portfolio weights for many markets are quite small ($< 0.001$), thus these markets will not be used for applications requesting $< 1000$ servers. As a result, the high risk-averseness scenario corresponds to about 10-15 server-types for most applications.

**Spark.** To see the impact of using application-specific policies, we run the Spark KMeans workload with the "medium" risk-averseness portfolio. We vary the fault-tolerance and replenishment policy, and show the increase in running time compared to on-demand servers when there is a market failure (corresponding to about 1/5th of the servers lost) in Table 1. The periodic checkpointing with eager replenishment increases job length by 7% compared to on-demand servers, while the no-checkpointing, no-replenishment policy increases job length by 54%. Overall, these policies yield 60-80% cost savings compared to on-demand servers, while increasing completion times by 1.07-1.5×.

Next, we observe Spark performance as a function of portfolio diversity and fault-tolerance policy, and use eager replenishment. Figure 9 shows the impact of a single revocation event on the running time (compared to on-demand servers). Since checkpointing reduces RDD recomputation in Spark, the increase in job completion times is only 5% in highly diversified portfolios and 16% in the low risk averseness portfolio. Even without checkpointing, using portfolios yields 62%, 38%, and 18% increases in running time with low, medium, and high diversification, while still yielding 86% cost savings.

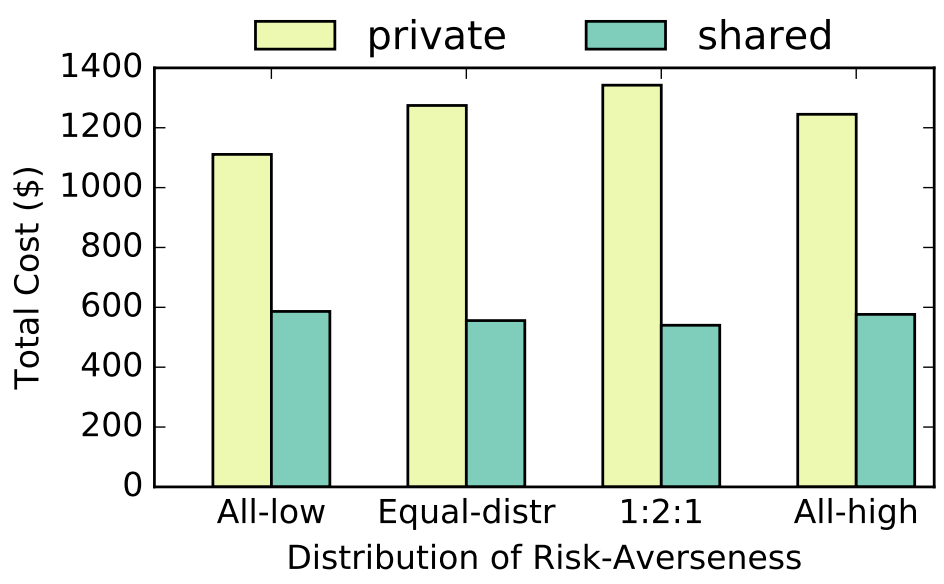

Fig. 11. Packing policy comparison.

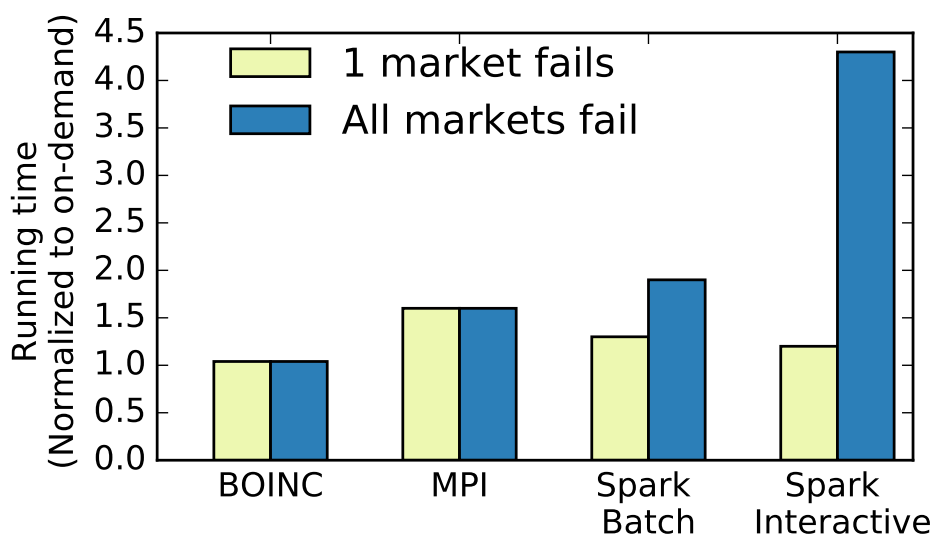

Fig. 12. Performance during the worst-case "black swan" failure events.

As the portfolio diversity increases, the impact of revocation of servers in one market decreases, because a smaller fraction of the state is lost. The recovery of the lost state is the primary contributor to the increase in running time. Thus, Spark workloads benefit from portfolio diversification, and can reduce the increase in running times to 5% using checkpointing, and 86% cost savings compared to on-demand servers.

**MPI.** For MPI, we examine the *expected* running time of jobs compared to on-demand servers in Figure 10. We consider two cases. In the first case, we assume that the job-lengths are known in advance, and only request markets with MTTRs greater than two times the job-length. When only a single-market portfolio is chosen ("MPI-MTTR" in the figure), the running time increases by only 3% compared to on-demand servers. In the second case, we use the Spot Fleet strategy of picking the cheapest single market. In this case ("MPI-Cost" in the figure), the running time increases by 10%. Thus, picking stable markets with MTTR >> job-length significantly reduces the running time for MPI, when compared to the lowest-cost strategy. The cost-savings are ~80% in all cases.

Note that since a single server revocation stops the entire MPI job, MPI only cares about minimizing server revocations, and *not* simultaneous revocations. Portfolio diversification reduces the number of simultaneous revocations and not the total number of revocations. The effect of increasing portfolio diversity can also be seen with the "MPI-MTTR" strategy in Figure 10, and we see that for highly diversified portfolios, the increase in running-time is close to 10%. Thus, MPI does not benefit from portfolio-diversification, and stable, single-portfolios represent the best portfolio.

**BOINC.** Unlike MPI, revocation in BOINC only affect the tasks that are running on the servers. We show BOINC performance compared to on-demand servers with the lazy-checkpointing and the restart fault-tolerance policies in Figure 10. With the highly diversified portfolio, the increase in running time compared to on-demand servers is less than 1% with the lazy-checkpointing and about 2% without checkpointing. Thus, ExoSphere can run embarrassingly parallel tasks at 1.01-1.05× overhead, but at more than 80% cost savings.

**Result:** *The impact of portfolio diversification is highly application dependent. Application performance is governed by the combination of portfolio composition, application characteristics, fault-tolerance policy, and recovery policy. Cost depends on application performance and a combination of portfolio composition and application policies. Portfolios with low risk averseness (i.e. high risk portfolios) yield lower costs, albeit at higher revocation risk.*

## 6.3 Multiple Applications

ExoSphere is designed to run multiple applications simultaneously, and we now evaluate its multiplexing capabilities. We use the Eucalyptus cloud traces [4] for realistic application arrivals and resource requests. We evaluate the costs of running such a trace by using a simulator which replays the spot price traces. The purpose of our simulator is to observe cluster utilization and costs for different application usage scenarios and cloud prices.

Sharing of servers among multiple applications can reduce total costs. We evaluate the extent of these savings compared to not sharing servers (e.g., a private mode). We assign the risk-averseness to the jobs in the Eucalyptus trace using four different distributions: all jobs requiring low risk-averseness portfolios; all high risk-averseness portfolios; equally distributed among low-medium-high; and distributed in a 1:2:1 ratio. Figure 11 shows the total cost incurred with the private and the cluster sharing policy, which uses first-fit bin-packing to find idle resources in servers in each market to satisfy portfolio requirements. By sharing servers among multiple applications, this packing policy lowers costs by 50%.

### 6.4 Black Swans: Multiple market failures

Finally, it is important to note that while the portfolio construction technique gives the best expected portfolios, it assumes that the historical price trends will continue to hold. In extreme situations, it is possible that even when selecting servers with low risk of concurrent revocations, all (or a large majority) of markets might be revoked. These events are akin to stock market crashes and are the black-swan events that have a high impact and are hard to predict. We show the performance implications of such extreme events in Figure 12, which shows the relative performance of applications running on their ideal portfolios and using the "best" fault-tolerance policy. We compare the application performance in the expected case of a single-market failure versus the worst case when all markets fail. We see that the impact on different applications is varied. BOINC and MPI see no difference in their expected and worst-case, since their preferred portfolios have only a single market. For a batch workload in Spark, the increase is significant (50%), and for the interactive Spark workload, the increase is more then $4x$. We note that these black-swan events only cause a one-time performance-hit, and don't affect expected cost savings.

**Discussion:** The real-world success of any portfolio-based technique relies on the ability to model the returns and risks of the underlying markets. However, there are many events that cannot be modeled using historical price traces alone. Black-swan and other rare events are hard to model, since they may have never occurred in the past. We also note that transient instances can be unilaterally revoked by the cloud provider, and cannot be modeled by price-traces alone. While price-based modeling has led to great gains in financial markets, spot markets are different from classic financial markets in a number of ways. Spot prices show higher volatility—prices can increase 10X in a single jump. The high volatility makes spot markets harder to model and also means that existing financial models that assume low volatility cannot be applied directly.

## 7 RELATED WORK

Our work leverages prior work on transient servers, cluster management, and portfolio theory.

**Systems for Transient Servers.** Recent work has looked at developing systems and middleware for transient servers like EC2 spot instances. SpotCheck [45] introduced the notion of a *derivative cloud* which combines spot and on-demand instances to run arbitrary applications on top of spot instances with high availability. SpotCheck relies on nested virtualization and continuous memory checkpointing to live-migrate to on-demand instances upon revocation. For batch jobs, SpotOn [49] performs spot market selection by considering the market cost and availability, and showed that the fault tolerance mechanism has an important influence on the server selection. OptiSpot [25] uses a combination of queueing-based application performance models and a markov chain based spot price models to select the right server type and bid price for a given application.

**Transient server selection.** ExoSphere's portfolio based server selection differs from prior work in regards to its flexibility and generality. The risk tolerance knob introduced in ExoSphere allows easy and explicit characterization of portfolio risk. Transient server selection policies in earlier systems [20, 36, 43, 45, 49, 57] do not have explicit support for managing revocation risk. This is because these systems have mostly targeted a single class of applications, and have server selection policies suited to that. For example, Flint [43] runs Spark [61] applications on transient cloud servers, and selects markets with the lowest effective cost for batch Spark jobs, and uses a greedy multi-market strategy for batch-interactive jobs.

**Transiency-aware Applications.** Prior work on making applications transiency-aware has involved application-level application-specific approaches. For example, Flint [43] and TR-Spark [57] modify Spark to better support transiency, e.g., via checkpointing, while related work focuses on optimizing MPI jobs for spot servers [36]. Similar work has modified Hadoop and other batch applications for transient servers as well [56, 59]. Checkpointing and scheduling policies for data processing and machine learning workloads on transient resources have been developed more recently in [31, 58].

**Portfolios.** In contrast to prior work which focused on supporting narrow classes of applications on transient servers, ExoSphere's goal is to provide a common platform for a wide range of applications. ExoSphere distills common abstractions based on the experience of past work to enable easy modification of current and future applications to support transiency. Our portfolio abstraction is inspired from financial economics, where investment portfolios are created for diversification and to reduce risk [26, 28, 37, 38] . In transient server markets, diversification reduces the probability of simultaneous revocations, and thus plays a crucial role in server selection. The idea of risk-reduction using diversification formalized in Modern Portfolio Theory in the 1950's remain the basis for other popular portfolio creation techniques such as Black-Litterman [40]. Exploring other portfolio construction techniques for server selection remains part of our future work.

A significant amount of prior work has gone into optimizing multiple objectives in the context of server selection. Server selection to optimize for performance and cost of on-demand servers (but without transiency considerations) is discussed in [29, 55], and in [53] which uses genetic algorithms to find spot/on-demand pareto-efficient frontier. CherryPick [11] uses bayesian optimization to select cost-optimal on-demand servers. Utility-based selection of servers is shown in [19], which selects homogeneous cloud servers for different applications. In contrast, ExoSphere selects a heterogeneous mix. Our use of portfolios is not to be confused with portfolios of policies/algorithms—wherein a portfolio of multiple policies and algorithms are run to find the most efficient algorithm. This approach is commonly used in SAT solvers [33], and has also been applied to cloud scheduling [48].

**Cluster Management.** There has been a significant amount of work on designing cluster resource managers [15, 32, 41, 51, 52] and resource management policies for running multiple applications in data center [22, 30, 34, 35] and cloud environments [23, 24]. In particular, ExoSphere builds on Mesos [32] and can be viewed as a transiency-aware cluster manager. To our knowledge, current cluster managers do not support transiency and variable pricing as first-class primitives.

**Resource Allocation in Data Centers.** There has also been a significant amount of work in allocation of surplus resources and risk-driven resource allocation in data centers. [18] allows idle resources to be reclaimed and uses resource usage traces to predict resource availability for long running services. Services can run uninterrupted with a high probability by maintaining slack between the resource allocation and usage. Risk-aware overbooking of resources by using admission-control policies is discussed in [50]. [42] considers job task placement to mitigate correlated failures in the data center, where a failure in a power component can affect multiple machines and hence multiple tasks. Using surplus resources in computational clusters has a long history—Spawn [54] introduced a market based system for selling idle resources to applications, similar to what public cloud operators are doing now. ExoSphere's portfolio-driven allocation policies can work in data center environments where the resource allocation involves optimizing two different and possibly competing objectives. For example, ExoSphere can be used to minimize performance interference adjusted costs, where instead of revocation risk, there is risk of performance interference due to co-located applications. We note that the transiency specific API that we have developed for applications can be used "as-is" in data center environments, where the resources of low priority jobs are revoked in favour of higher priority applications.

## 8 CONCLUSION

The effective use of transient servers is predicated on their careful selection. In this paper, we introduced portfolio modeling for transient server resource management. Unlike prior resource management schemes, portfolios allow the easy creation of virtual clusters with different revocation risk tolerances. Existing convex optimization techniques can compute portfolios efficiently—computing portfolios for 2,500 spot markets takes well under one minute.

We have prototyped and implemented a portfolio-driven cluster manager, ExoSphere, that exposes a narrow, uniform interface and allows multiple applications to develop and use their own transiency-aware policies for handling revocations. We have shown that existing applications such as MPI, Spark, etc., can use this interface to design their own policies and significantly increase their performance and cost saving on transient servers. Our experience with portfolios has shown that they are a powerful and promising resource management primitive, and can be especially useful in situations where multiple resource management objectives (such as cost and revocation risk) have to be minimized.

**Acknowledgments.** We thank our shepherd Giuliano Casale and all the reviewers for their insightful comments that helped us improve the paper. This work is supported in part by NSF grants #1422245 and #1229059, a Google faculty research award, and by Amazon Web Services (AWS) Cloud Credits. Any opinions, findings, and conclusions or recommendations expressed in this material are those of the author(s) and do not necessarily reflect the views of the National Science Foundation.